# Estimation of speed, armature temperature and resistance in brushed DC machines using a CFNN based on BFGS BP

Hacene MELLAH[1,2*], Kamel Eddine HEMSAS[1], Rachid TALEB[2], Carlo CECATI[3]
[1] Electrical Engineering Department, Faculty of Technology, Ferhat Abbas Sétif 1 University, Sétif, Algeria.
[2] Electrical Engineering Department, Faculty of Technology, Hassiba Benbouali University, LGEER Laboratory, Chlef, Algeria.
[3] Department of Information Engineering, Computer Science and Mathematics, University of L'Aquila, Italy.
*Correspondence: h.Mellah@univ-chlef.dz

**Abstract:** In this paper, a sensorless speed and armature resistance and temperature estimator for Brushed (B) DC machines is proposed, based on a Cascade-Forward Neural Network (CFNN) and Quasi-Newton BFGS backpropagation (BP). Since we wish to avoid the use of a thermal sensor, a thermal model is needed to estimate the temperature of the BDC machine. Previous studies propose either non-intelligent estimators which depend on the model, such as the Extended Kalman Filter (EKF) and Luenberger's observer, or estimators which do not estimate the speed, temperature and resistance simultaneously. The proposed method has been verified both by simulation and by comparison with the measurements and simulation results available in the literature.

## 1. Introduction

In the last few years there has been growing interest in thermal aspects of electrical machines and their effects on electrical and mechanical parameters and time constants, such as electrical resistance, back EMF, and so on [1], since due to their influence, the motor's characteristics, and hence its performance, during operation are not the same as those considered during design [2]. Real-time knowledge of temperature in the various motor parts is also very useful in order to predict incipient failures and to adopt corrective actions, thus obtaining not only better control but higher reliability of the electrical machine.

In fact, the early prediction of thermal aging, which makes insulations vulnerable, as well as of other thermal factors directly influencing motor health and life can avoid dangerous failures [3-5].

The main causes of thermal faults are: overloads [6], cyclic mode [7], over voltage and/or voltage unbalances [8], distortions [4], thermal insulation aging [3], obstructed or impaired cooling [9], poor design and manufacture [3], and skin effect [10].

For several years, great efforts have been devoted to the temperature and speed measurement of electrical machines, and several methods for temperature [11-13] and speed measurements [14] have already been proposed in the literature. While the direct measurement of temperature in electric DC machines is a long-established approach [13-15], some authors obtained the average winding temperature from the resistance measurement [13]. A more modern method can be found in [12,16], but the temperature measurement gave rise to two major problems: optimum sensor placement and the difficulty of achieving rotor thermal measurements. Likewise, speed measurement can also be difficult [17]. Moreover, information from sensors installed on rotating parts leads to techno-economic difficulties in the measurement chain. Sensorless solutions have therefore been considered by many studies [16,18-20].

One of the first examples of temperature estimation is presented in [21], where a Luenberger observer was applied both to a DC rolling mill motor and a squirrel cage induction motor. Another solution was described in [22] where the authors used a steady-state Extended Kalman Filter (EKF) associated with its transient version. Nevertheless, to estimate the resistance some authors combine EKF with a smooth variable structure filter [23]. Some research on bi-estimation has been done [24], which describes and implements an algorithm for combined flux-linkage and position estimation for PM motors based on the machine's characteristic curves. A very interesting approach was proposed in [25], applying and experimentally validating a transient EKF to estimate the speed and armature temperature in a BDC motor. However, EKF has some limitations, in particular: (i) if the system is incorrectly modelled the filter may quickly diverge; (ii) the EKF assumes Gaussian noise [26-28]; (iii) if the initial state estimate values are incorrect the filter may also diverge; (iv) the EKF can be difficult to stabilize due to the sensitivity of the covariance matrices [27,29].

To the authors' knowledge, very few publications deal with the simultaneous estimation of speed and armature temperature of DC machines [25], especially when performed by intelligent techniques [29]. Artificial neural networks (ANN) have demonstrated their ability in a wide variety of applications such as process control [30], identification [31], diagnostics [32], pattern recognition [33], robot vision [34], flight scheduling [35], finance and economics [36] and medical diagnosis [37].

In this paper, while referring to our previous study [29], in which an estimator based on Multilayer Perceptron with Levenberg-Marquardt BP was developed in order to avoid the limitations of the standard ANN, a solution based on a Cascade-Forward Neural Network (CFNN) and Bayesian Regulation BP (BRBP) is proposed. A highly accurate BRBP-based ANN was proposed in [38,39] but it requires an enormous convergence time, and is in fact known to be among the slowest algorithms to converge.

Based on the approach already presented in [29], the purpose of this paper is to propose a novel approach using a learning algorithm which is a compromise between speed and accuracy. The BFGS can respond to these two constraints [39].

The remainder of the paper is organized as follows: Section II describes the thermal model of the BDC motor, Section III discusses ANN and CFNN based on Quasi-Newton BFGS BP, and Section IV presents the simulation results and analysis. Finally, some conclusions are discussed in Section V.

## 2. Thermal model of BDC machines

Research interest in studying rotating electric machinery from the combined viewpoints of thermal and electrical processes dates back to the 1950s [40,41]. The model used in the present paper was proposed by [25], and the thermal model is derived by considering the power dissipation and heat transfer [25]. The power is dissipated by the armature current flowing through the armature resistance, which varies in proportion to the temperature. The electrical equation can be written as:

$$v_a = R_{a0}(1 + \alpha_{cu}\theta)i_a + L_a \frac{di_a}{dt} + k_e \omega \qquad (1)$$

where $V_a$ is the armature voltage, $R_{a0}$ is the armature resistance at ambient temperature, $\alpha_{cu}$ ($\alpha_{cu} = 0.004$ /°C) is the temperature coefficient of resistance, $\theta$ the temperature above ambient, $i_a$ the armature current, $l_a$ the armature inductance, $k_e$ the torque constant, and ω the armature speed.

The electrical and mechanical behavior of the motor are coupled by the following equation:

$$J \frac{d\omega}{dt} + b\omega + T_l = k_e i_a \qquad (2)$$

where $J$ is total inertia, $b$ is the viscous friction constant and $T_l$ is the load torque.

The iron loss is proportional to speed squared for constant excitation multiplied by the iron loss constant $k_{ir}$ ($k_{ir} = 0.0041$ W/(rad/s)$^2$). The power losses $P_l$ include contributions from copper losses and iron losses which are frequency dependent:

$$P_l = R_{a0}(1 + \alpha_{cu}\theta)i_a^2 + k_{ir}\omega^2 \qquad (3)$$

Heat flow from the armature surface of the BDC motor is directly to the cooling air and depends on the thermal transfer coefficients at zero speed $k_o$ ($k_o = 4.33$ W/°C) and at $k_T$ ($k_T = 0.0028$ rad/s); The thermal power flow from the armature surface of the BDC motor surface is proportional to the temperature difference between the motor and the ambient temperature. The rate of temperature variation depends on the thermal capacity $H$ ($H = 18$ KJ/°C):

$$P_l = k_0(1 + k_T \omega)\theta + H \frac{d\theta}{dt} \qquad (4)$$

By arranging the previous Eqs., we can write the system of equations as:

$$\frac{di_a}{dt} = -\frac{R_{a0}(1+\alpha_{cu}\theta)}{L_a}i_a - \frac{k_e}{L_a}\omega + \frac{1}{L_a}v_a$$
$$\frac{d\omega}{dt} = \frac{k_e}{J} - \frac{b}{J}\omega - \frac{K}{J}T_l \quad (5)$$
$$\frac{d\theta}{dt} = \frac{R_{a0}(1+\alpha_{cu}\theta)}{H}i_a^2 + \frac{k_{ir}}{H}\omega^2 - \frac{k_0(1+k_T\omega)}{H}\theta$$

## 3. ANN estimator

In recent years, CFNNs have become very popular [42-51], and have proved their capability in several applications [49-57], becoming the preferred choice in [57]. Many authors [49-56] consider that CFNN are similar to feed-forward neural networks (FFNN), but include a weight connection from the input to each layer and from each layer to the successive layers. For example, a four-layer network has connections from layer 1 to layer 2, layer 2 to layer 3, layer 3 to layer 4, layer 1 to layer 3, layer 1 to layer 4 and layer 2 to layer 4. In addition, the four-layer network also has connections between input and all layers. FFNN and CFNN can potentially learn any input-output relationship, but CFNNs with more layers might learn complex relationships more quickly [50-53], making them the right choice for accelerated learning in ANNs [51]. The results obtained by Filik et al. in [52] suggest that CFNN BP can be more effective than FFNN BP in some cases.

In the proposed application, CFNN voltage and current are inputs and speed, while armature temperature and resistance are outputs. The performance and robustness of the CFNN was tracked by adding random white Gaussian noise to inputs, as shown in Figure 1. The BP algorithm was used to form the neural network such that on all training patterns, the sum squared error '$E$' between the actual network outputs, '$y$', and the corresponding desired outputs, $y_d$, is minimized to a supposed value:

$$E = \sum(y_d - y)^2 \quad (6)$$

To obtain the optimal network architecture, for each layer the transfer function types must be determined by a trial and error method. On the input and hidden layer, a hyperbolic tangent sigmoid transfer function was used, defined as:

$$f(net_j) = \frac{2}{1+e^{-2net_j}} - 1 \quad (7)$$

where net is the weighted sum of the input unit, and *f(net)* is the output units. The output layer has 3 units with a pure linear transfer function, defined as:

$$f(net_j) = net_j \tag{8}$$

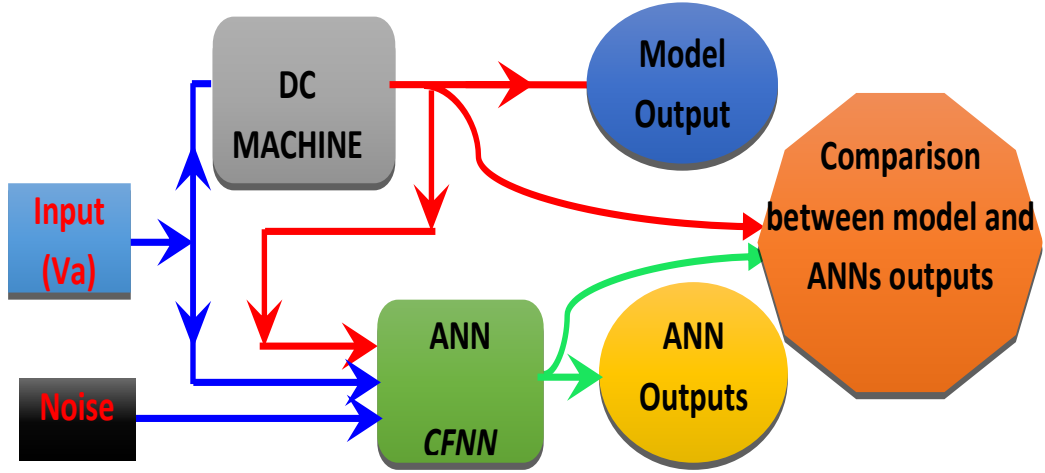

**Figure 1.** Comparison between model and ANN outputs

## 3.1. Quasi-Newton BFGS BP algorithm

The Quasi-Newton BFGS BP training algorithm is a useful method for updating network weights and biases according to the BFGS formulae [58-61]. The algorithm belongs to the quasi-Newton family and was devised by Broyden, Fletcher, Goldfarb, and Shanno in 1970 [62-65] to achieve fast optimization [60-61]. It is an iterative method that approximates Newton's method without the inverse of Hessian's matrix [60]. It is a second order optimization algorithm [60-61]. In this paper, the weight and bias values were updated according to the BFGS quasi-Newton method, and the new weight $w_{k+1}$ was computed as:

$$w_{k+1} = \omega_k - H_k^{-1}\Psi_k \tag{9}$$

where: $H_k$ is the Hessian matrix of the performance index at the current values of the weights and biases. When $H_k$ is large, $w_{k+1}$ computation is complex and time consuming [66-68]. BFGS do not calculate the inverse Hessian but approximate it as follows:

$$H_{k+1} = H_k + \frac{y_k y_k^T}{y_k^T s_k} - \frac{H_k s_k s_k^T H_k}{s_k^T H_k s_k} \tag{10}$$

where: $\Psi_k = \nabla f(w_{k+1})$, $S_k = w_{k+1} - \omega_k$ and $y_k = \nabla f(w_{k+1}) - \nabla f(w_k)$. The new formula can be approximated as:

$$w_{k+1} = \omega_k - \left(H_k + \frac{y_k y_k^T}{y_k^T S_k} - \frac{H_k S_k S_k^T H_k}{S_k^T H_k S_k}\right) \Psi_k \qquad (11)$$

This method has several advantages: it has a better convergence rate than using conjugate gradients [58-61], it is stable because the BFGS Hessian update is symmetric and positive definite [60]; in addition, BFGS computes an approximation to the inverse Hessian in only $O(n^2)$ operations [60]. However, this method requires a lot of memory to converge, especially on a large scale [66-69], whereas many researchers are interested in how to reduce memory needs [67-71].

**4. Simulation results**

The estimated speed, armature temperature and resistance are shown in Figures 2 to 5 for a continuous running duty or abbreviated by duty type S1. Duty type S1 is characterized by operation at a constant load maintained for a sufficient time to allow the machine to reach thermal equilibrium [72]. The ANN outputs are in good agreement with the model outputs as can be seen below, proving the ability of the proposed approach. The BDC motor parameters used during simulations are as follows:

- Rated voltage $V_a$ = 240 V

- Rated power $P$ = 3 kW

- Rated torque $T_l$ = 11 Nm

- Armature resistance $R_{a0}$ = 3.5 Ω

− Armature inductance $L_a$ = 34 mH.

The estimated speed and the corresponding errors are shown in Figure 2. The results obtained by Acarnley et al. in [25] suggest that the speed estimation error from EKF is approximately 2%. Moreover, it is not suitable for high-performance servo drives [25]. However, in the results obtained here, the error is less than 0.4 rpm and represents only 0.18% of the final value, as shown in Figure 5.

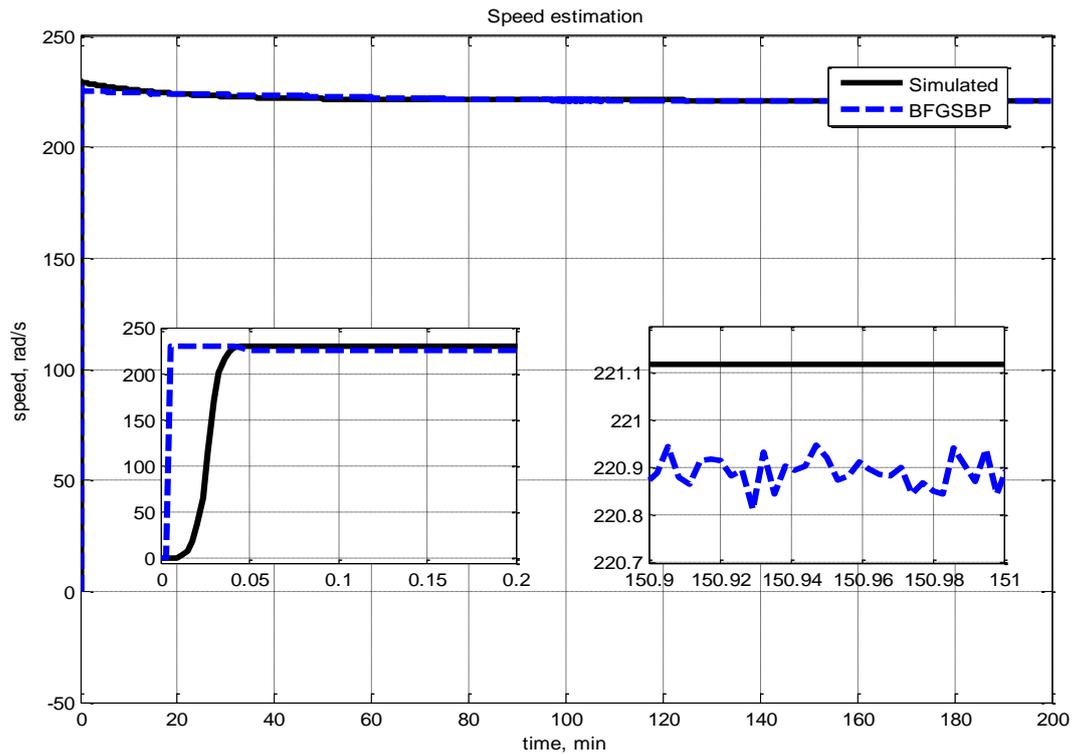

**Figure 2.** Estimated and simulated speed.

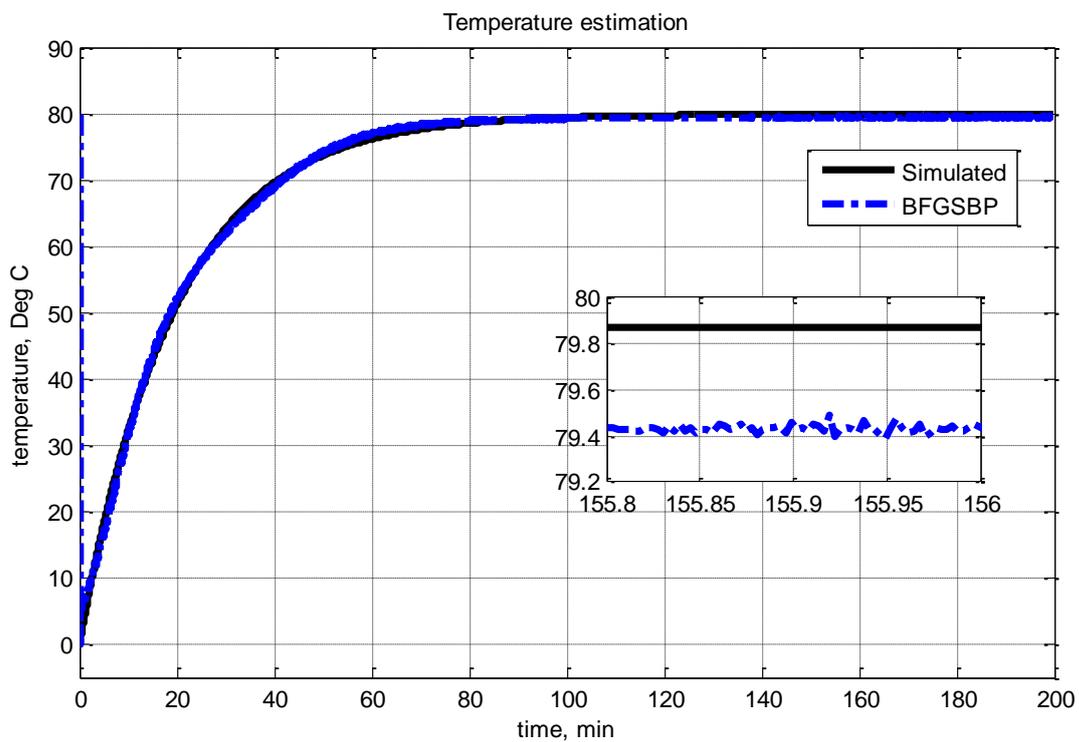

**Figure 3.** Estimated and simulated armature temperature.

The estimated temperature and the corresponding errors are shown in Figure 3 where it reaches 79.5 °C, while the model output is 80 °C; the steady state estimated error is less than 0.5 °C as can be seen from Figure 5. This can be contrasted with the results in [25] which suggest that the temperature estimation error from EKF is 3 °C, i.e. approximately 3.75%, while Nestler et al. in [21] using a Luenberger's observer found that the estimated winding temperature error was high. The results shown in [22], however, suggest that the error is not greater than 1 °C and the results presented in this paper show that the error is insignificant (0.5 °C) and represents only 0.625% as can be seen from Figures 3 and 5.

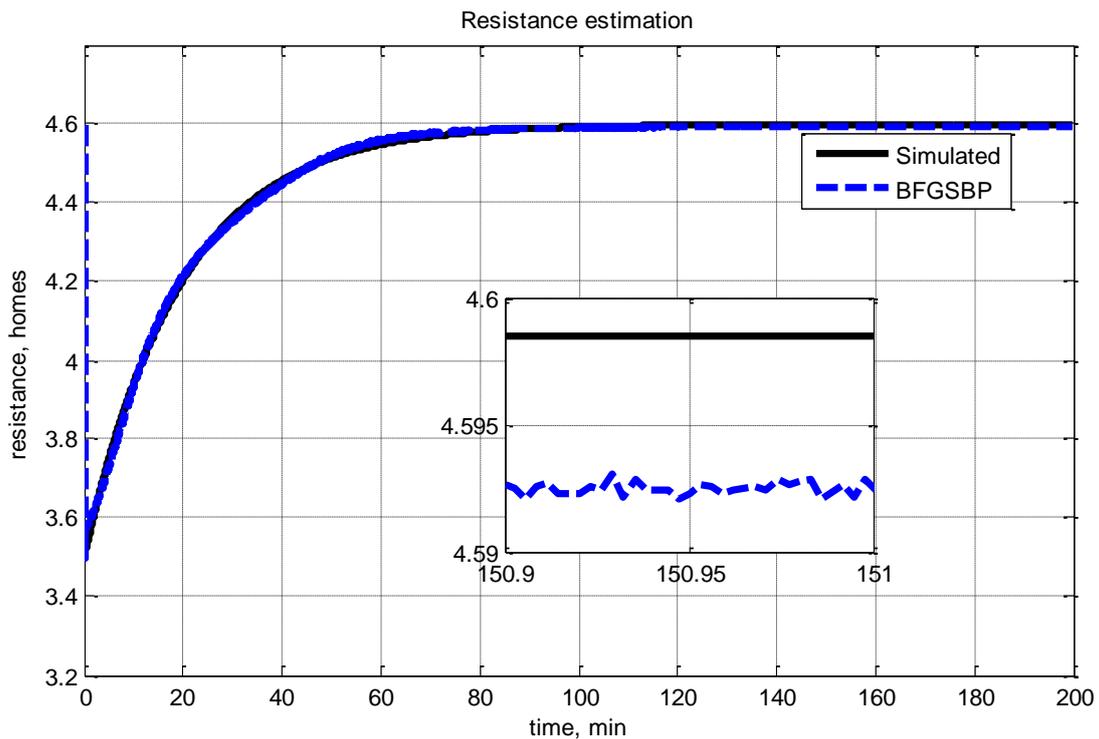

**Figure 4**. Estimated and simulated armature resistance.

Figure 4 depicts the resistance estimated by ANN and the model response. From this Figure, it can be seen that the resistance has the same curvature as the armature temperature, where the steady state estimated resistance is 4.59 Ω, i.e. less than $6 \times 10^{-3}$ of the simulated resistance. Practically, this difference is a negligible quantity and represents only 0.13% of the final value. The results obtained are more precise than those presented in [23]. Figure 5 shows the estimation errors of speed, temperature and resistance, and their percentage in relation to their rated value. This Figure shows more clearly the excellent agreement between the model outputs and the outputs of our intelligent sensor.

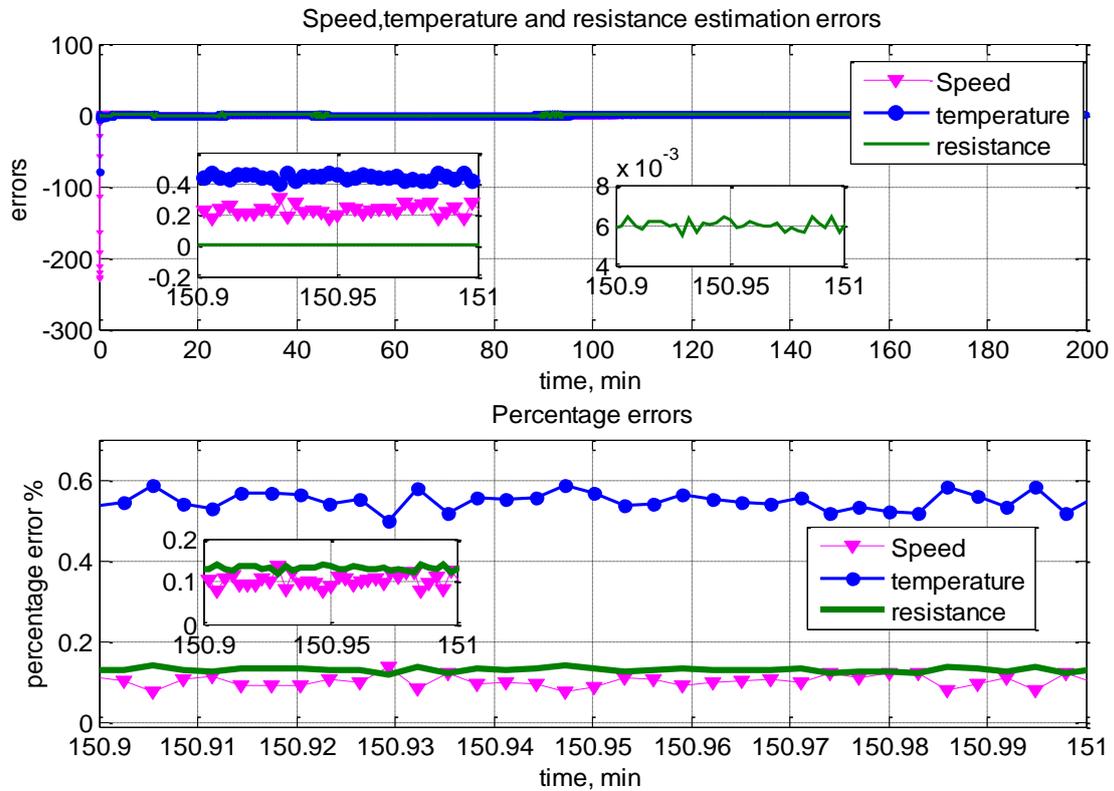

**Figure 5**. Speed, temperature and resistance estimation errors

## 5. Conclusion

A sensorless speed and armature winding quantity estimator has been proposed for BDC machines based on a CFNN trained by BFGS BP. The estimator includes sensorless speed estimation, average armature temperature and resistance estimations based only on the voltage and the current measurements. Estimated speed and temperature eliminate the need for speed measurements and the need for a thermal sensor. In addition, estimated temperature solves the problem of obtaining thermal information from the rotating armature. Furthermore, the estimated resistance can be used to improve the accuracy of the control algorithms which are affected by an increase in resistance as a function of temperature. The good agreement between the model and the intelligent estimator demonstrates the efficiency of the proposed approach.